# Ultralong Photon Echo Storage Using Optical Locking

Byoung S. Ham[1,2,3,*] & Joonseong Hahn[1,2]

[1]Center for Photon Information Processing,
[2]The Graduate School of Information and Telecommunications, and
[3]Department of Electrical Engineering, Inha University
253 Yonghyun-dong, Nam-gu, Incheon 402-751, S. Korea
[*]bham@inha.ac.kr

An ultralong photon storage with more than 50% retrieval efficiency is presented in the three-pulse photon echoes by using a pair of control pulses, where the control pulses play to freeze atoms' dephasing process conditionally by transferring excited atoms to an auxiliary spin state and returning them back later. The observed photon echo decay time much longer than the spin homogeneous decay time is potential for long distance quantum communications.
PACS numbers: 42.50.Md (Optical transient phenomenon, photon echo), 78.47.jf (photon echoes),

Recently photon echoes[1,2] have been intensively studied for quantum memory applications owing to the benefit of multimode storage in the time domain as well as wide bandwidth processing using reversible inhomogeneous broadening[3-13]. However, most modified photon echo protocols[5-11] are still limited by a short storage time confined by spin phase relaxation. Conventional stimulated photon echoes with the advantage of extended storage time may not be applied directly into quantum memories due to their intrinsically low echo efficiency[2]. Here, we report an ultralong photon storage in controlled stimulated photon echoes, where the storage time extends far beyond the conventional limitation due to a frozen phase via optical locking. We have performed experiments in a rare-earth doped solid and demonstrated that the storage time is extended by several orders of magnitude, where echo signals saturate at 50% of maximum retrieval efficiency even far beyond the spin homogeneous decay time.

Quantum optical data storage is a key element in quantum information processing such as quantum computing and long-haul quantum communications based on quantum repeaters[14,15]. As a classical counterpart of all-optical memories, quantum memories are also characterized by the following criteria: access rate, bandwidth, and storage time. Even though slow light-based quantum memories utilizing a quantum mapping process have been demonstrated using nonclassical light, a single mode feature with limited storage time may limit its effectiveness for practical quantum information processing[16-19]. Photon echoes[1,2], on the other hand, have been intensely studied for possible quantum memory applications owing to the high speed access rate and ultrawide bandwidth[3-13]. Recently the Geneva group demonstrated photon echo-based quantum optical data storage using single photons in a rare-earth doped solid medium. Although the observed retrieval efficiency was extremely low at ~1% (ref. 3), photon storage time should be beyond the spin dephasing limitation because it is basically an ultimate case of three-pulse photon echoes[4]. Quantum memories may require more than 50% retrieval efficiency in amplitude. In this context, conventional two-pulse photon echoes basically satisfy the requirement of quantum memories. However, the storage time in two-pulse photon echoes is short, where optical phase decay time is a practical constraint (~100 μs in rare earth doped solids)[9]. Spontaneous emission noise from the π rephasing pulse is another obstacle to overcome[20]. Unlike the two-pulse photon echoes, three-pulse (or stimulated) photon echoes offer much longer storage time, even though they may not fulfill the quantum memory requirement due to less than 50% retrieval efficiency caused by optical population loss during the storage process[2].

Over the last several years, photon echo studies have sought to satisfy quantum memory applications in modified schemes: 1. to increase echo efficiency[4-8,21], 2. to remove spontaneous emission noise[3,4,6,7,9-11], and 3. to extend storage time[4,6-8,11-13]. For example, storage time extension has been achieved in a modified two-pulse photon echoes[6,8]. However, the storage time extension is limited by the spin dephasing process, a mere fractional delay (if spin inhomogeneous broadening is involved) compared with the constraint of optical homogeneous (phase) decay time[6-8]. The spin population relaxation-limited photon storage method utilizing spin inhomogeneous broadening via resonant Raman transitions is the first quantum version of general three-pulse photon echoes[13]. Off resonant Raman echoes utilizing controlled spin inhomogeneous broadening via magnetic field gradient have been experimentally performed in an atomic vapor, where inversing magnetic field plays a rephasing process[11]. The long preparation stage[3,4], use of magnetic field gradient[11], or spin-optical conversion process[13,22], however, may be a practical drawback to implement the modified photon echoes.

Storage time extension as well as higher retrieval efficiency has been an urgent task in the photon echo studies performed so far. Here, we report an optically locked echo protocol based on the three-pulse (stimulated) photon echoes for ultralong photon storage extended up to spin population decay time even in the presence of spin inhomogeneity, with more than 50% retrieval efficiency regardless of the storage time. Unlike previous methods in ultralong photon storage[4,13], the present scheme is much simpler and more practical using a pair of short optical pulses. The idea for the present scheme originates in the phase locked photon echoes. Ultralong photon storage in the present method is achieved by locking both individual phase evolution and population decay loss of the excited atoms. Numerical calculations of the photon storage matches quite well the observed data, where the echo decay time reaches far beyond the spin homogeneous decay time. Moreover, the observed echo signals gradually saturate on 50% of the maximum amplitude even far beyond the spin homogeneous decay time. Like optically controlled phase locked echoes[5,8], resonant Raman echoes[13], or gradient echoes using AC Stark field for rephasing process[6], the optical locking is achieved by transferring excited atoms to an auxiliary ground state and returning them back by using a pair of control lights. The individual atom phase locking is a direct result of the conventional three-pulse photon echo method. In the phase locked photon echoes, individual atom phase evolution is not actually locked, but neglected due to zero coherence resulting from evacuation of the excited atoms[8]. Thus, in the phase locked echo technique, the atom phase decay must be influenced by the spin inhomgeneous broadening because of transferred atoms.

      The motivation of the present research is to solve this spin inhomogeneity-based decay problem. As a result, all benefits of stimulated photon echoes in the classical regime are taken into the quantum regime satisfying ultralong photon storage with more than 50% retrieval efficiency. In most rare-earth doped solids[23], the spin phase relaxation time is orders of magnitude longer than either the optical population relaxation time (as a constraint of conventional stimulated photon echoes), or the spin dephasing time (as a constraint of phase locked echo)[23]. As presented already in other methods[4,13], the photon storage time of the present method is limited by spin population decay time, which is a minute or more[24]. In atomic media, the storage time extension by the present technique also should be several orders of magnitude longer[16].

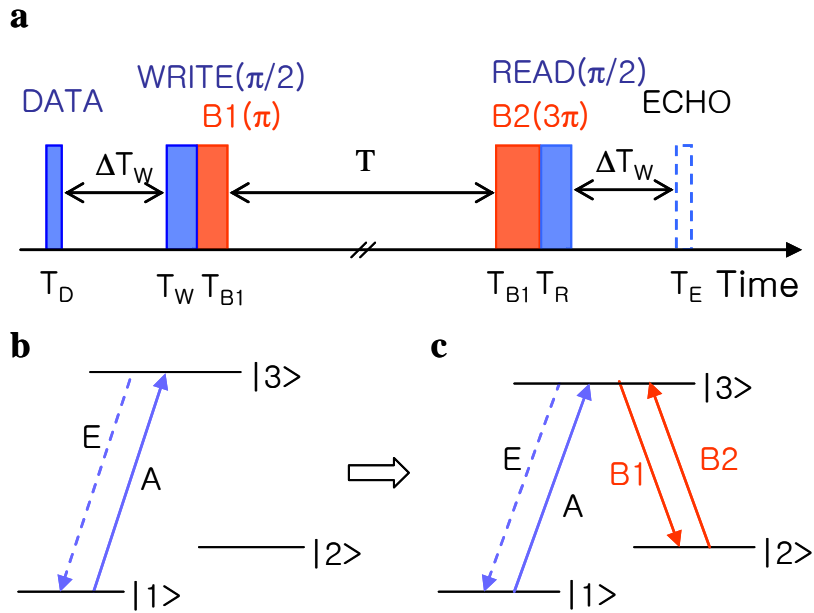

**Figure 1. Schematic of photon echoes using optical locking. a,** Pulse sequence. **b,** Energy level diagram for conventional photon echoes. **c,** Energy level diagram for (a).

Figure 1 shows a schematic of the present optically locked photon echoes. Figure 1a represents a block diagram of the pulse sequence for the present protocol: Parentheses stand for the pulse area $\Phi$: $\Phi = \int \Omega dt$, where $\Omega$ is the Rabi frequency of the applied field. Figure 1b shows a two-level system of conventional photon echoes. Fig. 1c shows the present scheme with control fields B1 and B2 and an auxiliary state |2>. Each colored arrow in Figs. 1b and 1c is color matched with Fig. 1a. When the state |2> is isolated from state |1>, it forms a perfect scheme for infinite storage time of quantum memory.

For the numerical calculations, nine time dependent density matrix equations under rotating wave approximations are solved. The equation of motion of the density matrix operator $\rho$ is determined from Schrödinger's equation:

$$\frac{d\rho}{dt} = -\frac{i}{\hbar}[H,\rho] - \frac{1}{2}\{\Gamma,\rho\}, \qquad (1)$$

where $\{\Gamma,\rho\}$ is $\Gamma\rho+\rho\Gamma$. For photon echoes, optical inhomogeneous broadening is required. Optical inhomogenity can be controlled by using spectral hole burning. We assumed Gaussian shaped optical inhomogeneous broadening of 680 kHz at full width at half maximum for the optical transitions. For the calculations, the 1.6 MHz inhomogeneous region is divided into 161 segments. The Rabi frequency of rephasing pulse R and W (B1 and B2) is set at 5 MHz (10 MHz). Each pulse area is controlled by adjusting the corresponding pulse duration. For visual effect the data pulse area is set at $\pi/2$, which reaches maximum coherence excitation.

Figure 2 shows numerical simulations for Fig. 1a. The blue (red) curve in Fig. 2a represents absorption, Im$\rho_{13}$ without (with) the control light B1 and B2, corresponding to the three-pulse photon echo (optically locked echo). The green curve is for two-pulse photon echo as a reference, where the WRITE ($\pi/2$) and READ ($\pi/2$) pulses in Fig. 1a are combined together for a $\pi$ pulse. As seen the optically locked echo (red curve) shows the same echo magnitude as the reference (green curve). Instead the three-pulse photon echo (blue curve) is severely reduced due to population decay loss. In the numerical calculations there is no any kind of assumption except no phase evolution in the period between the WRITE and READ pulses according to the photon echo theory. Figures 2b and 2c show corresponding absorption spectra for the inhomogeneously broadened atoms. Unlike Fig. 2b (for the blue curve in 2a), Fig. 2c (for the red curve in 2a) shows nearly perfect retrieval phenomenon, even though the storage time (80 µs) is much longer than optical population decay time $T_1$ ($T_1=1/\pi\Gamma=1/\pi20$kHz ~ 16 µs) (see also Supplementary Fig. S1). Figure 2d shows a time slot of Fig. 2c at t=85.4 µs for the echo: The cyan curve is given as a perfect echo. The wiggling in the echo (red curve) is from detuning dependent population difference between states |3> and |1> (see also Supplementary Fig. 2). Figure 2e shows a typical spectral grating generated from DATA and WRITE pulses (see the red curve at t=$T_W$=10.1 µs).

The function of the READ pulse (at t=$T_R$=50.4 μs) is in retrieving of the spectral information resulting in the echo in Fig. 2d (see the similar wiggled pattern in the blue curve).

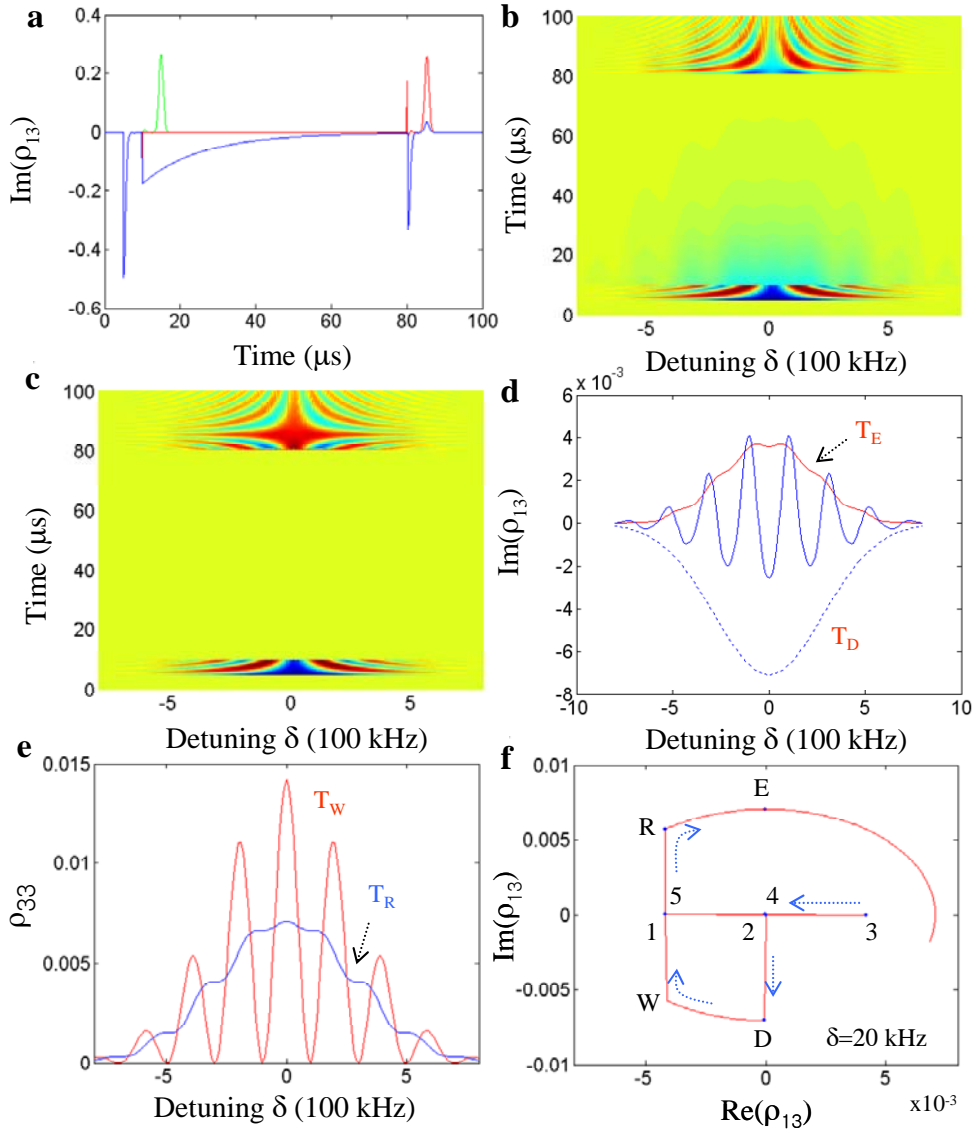

**Figure 2. Numerical calculations. a,** Photon echo simulations for conventional two-pulse scheme (green curve), three-pulse scheme (blue), and optically locked scheme (red): $\Gamma_{12}$=0; $\gamma_{12}$=0; $\Gamma_{31}$=$\Gamma_{32}$=10 kHz; $\gamma_{31}$=$\gamma_{32}$=10 kHz. $\Omega_D$=$\Omega_R$=2.5 MHz; $\Omega_{B1}$=$\Omega_{B2}$=5 MHz, where $\Omega\iota$ is Rabi frequency for light $\iota$. $\rho_{11}$(t=0)=1; $\rho_{22}$(t=0)=$\rho_{33}$(t=0)=0. **b,** Detuning δ versus absorption (Im $\rho_{13}$) for blue curve in a. **c,** Detuning δ versus absorption (Im $\rho_{13}$) for red curve in a. **d,** Absorption spectra of red curve in a. Cyan curve is for no decay. **e,** Excited population versus detuning of red curve in a. **f,** Bloch vector evolution for a detuned atom. All decay parameters assumed zero.

Figure 2f shows the optical locking mechanism in the Bloch vector uv plane with no decay for δ (20 kHz) detuned atoms. The coherence excited by DATA pulse builds up to D position and then decays freely with detuning δ. When the WRITE pulse (π/2 pulse area) switches on at W position, each atom's phase evolution should be frozen according to the photon echo theory, leading to point "1." Then the followed control pulse B1 (π pulse area) depopulates the excited atoms completely leading to point "2" generating a π/2 phase shift. The point "2" represents for a complete frozen phase state because of no population decay loss for the phase locked atoms. Although the second control pulse B2 can make a complete population transfer back to state |3> for a π pulse area, total accumulated phase shift π reverses the system reaching at point "3" (see refs. 8, 13). Therefore, an additional π phase shift is required. This is why the pulse area of the second control pulse B2 must be 3π. In other words, the pulse area of B1 and B2 together should satisfy 4nπ, where n is an integer. Unlike rephasing control in ref. 8, the optical phase evolution lock by W pulse should suspend the spin inhomogeneous broadening-dependent dephasing process, too. This is the key mechanism how ultralong storage can be obtained via optical locking in the present method.

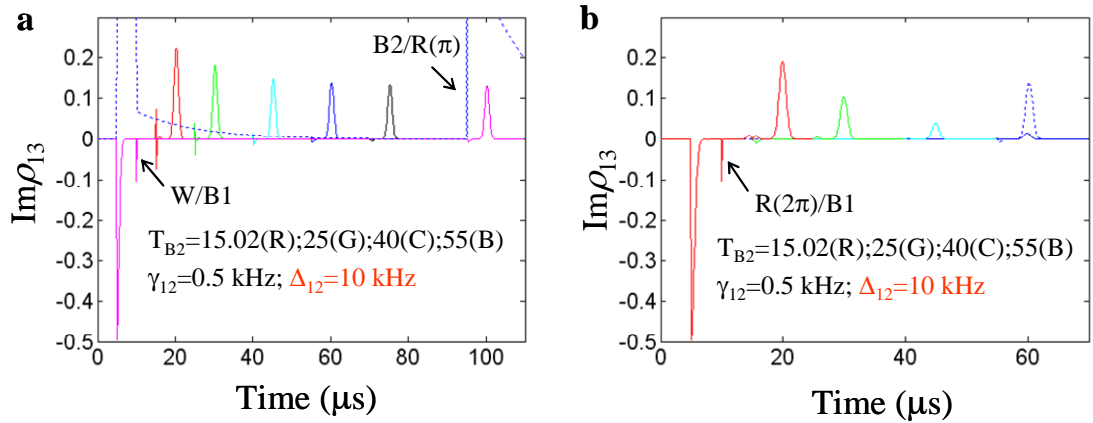

**Figure 3. Optically locked photon echoes with spin inhomogeneous broadening $\Delta_{12}$. a,** optically locked echo with $\Delta_{12}$=10 kHz. **b,** phase locked echo with $\Delta_{12}$=10 kHz. The dotted curve in (a) represents optical population decay $\rho_{33}$ for the magenta curve, where the decay causes the echo degradation. Dotted curve is for the blue in (a) for comparison. Parameters are the same as in Fig. 2 except $\Gamma_{31}=\Gamma_{32}$=5 kHz; $\gamma_{31}=\gamma_{32}$=10 kHz; $\gamma_{12}$=0.5 kHz; $T_{B2}$=10.02 (Red); 25 (Green); 40 (Cyan); and 55 (Blue) μs of Fig. 1a. $T_D$=5; $T_W$=10; $T_{B1}$=10.1 μs.

Figure 3 shows optically locked photon echoes affected by spin dephasing $\Delta_{12}$=10 kHz. The red, green, cyan, blue, black, and magenta curves are for different delay T in Fig. 1a, where T is the delay between B1 and B2 for $T_{B1}$=10.1 μs, and $T_{B2}$=10.2, 25, 40, 55, 70, and 95 μs, respectively. In each case the READ pulse follows B2 with no delay. In the simulations, we treated the spin dephasing induced by spin inhomogeneous broadening $\Delta_{12}$ as an effective spin phase decay $\gamma_{12}^{eff}$ in Fig. 3a because of the locked atom phase evolution by WRITE pulse. For the phase locked echo in Fig. 3b, however, the $\Delta_{12}$ should be treated as it is by giving relative dephsaing between optical $\Delta_{13}$ and $\Delta_{23}$: This treatment has been proved experimentally. Unlike the conventional two-pulse photon echoes in Fig. 3b, however, the optically locked echo survives much longer than $1/\gamma_{12}^{eff}$. An astonishing result is that the echo signal never drops down below 50% of the maximum value. This is definitely due to the optical locking via the control pulse set. Thus, the storage time extension is several orders of magnitude longer compared with the constraint of most modified photon echo protocols in rare-earth doped solids.

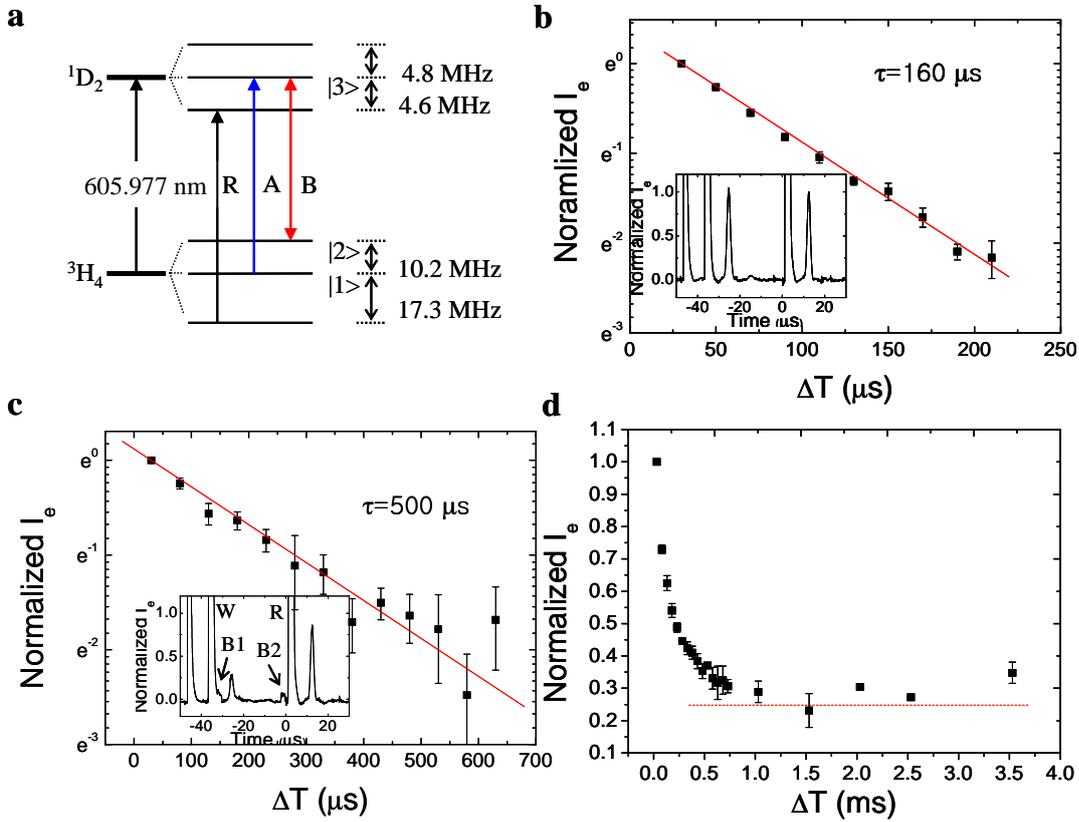

**Figure 4. Observations of optically locked echoes in Pr:YSO at T=5K. a.** Partial energy level diagram of Pr:YSO. R is used for repumping. Optical power for R, A, and B is 25, 4.6, and 14.5 mW, respectively. Opticalpulse length of D, W, R (B1 and B2) is 1 μs (1.2 μs). **b,** Stimulated photon echo decay time as a reference. **c,** Optically locked echo decay time. **d,** Echo saturation far beyond the spin phase decay time.

Figure 4 shows experimental proofs of the present proposal discussed in Figs. 1~3. The light pulses in Fig. 4a are from a ring-dye laser (Tecknoscan) via acoustooptic modulators (Isomet) driven by digital delay generators (Stanford DG 535) and radiofrequency synthesizer (PTS 250). The angle among beams A and B in Fig. 1 is 12.5 miliradians and overlapped by 90% through the sample of 1 mm in length. The light power of P, A, and B in Fig. 4a is 25, 4.6, and 14.5 mW, respectively. The optimum power of the light B1 and B2 is predetermined by Rabi flopping measurement, where the B1 (B2) pulse area in Fig. 1a is set at $\pi$ ($3\pi$). The optical signals detected by avalanche photodiodes are recorded in the oscilloscope by averaging 30 samples. The repetition rate of the light pulse sequence is 20 Hz. The temperature of the sample of Pr:YSO is controlled by the liquid helium evaporation rate in a cryostat (Advanced Research Systems). To utilize state $|2\rangle$ ($\pm 5/2$, $^3H_4$) in Fig. 4a as an auxiliary spin state for B1 and B2, both states except state $|1\rangle$ ($\pm 3/2$, $^3H_4$) are prepumped by B and P for about 1 ms to evacuate all atoms and to redistribute them to states $|1\rangle$ before sending the pulse chain in Fig. 1a. The effective atom broadening in state $|1\rangle$ is determined by laser jitter, which is ~300 kHz. All light beams are vertically polarized and nearly copropagate in a forward direction along the crystal (Pr:YSO) optical axis. The pulse D follows W at 10 µs delay, and the pulse B1 is followed by W at 5 µs delay. The delay between B2 and R is 3 µs. The pulse duration of D, W, R is 1 µs. The pulse duration of B1 (B2) is 1.2 µs (3.6 µs). The diameter of the light beam for D, W, and R at a focal point is 300 µs. The corresponding diameter of B for B1 and B2 is 330 µs.

Figure 4a shows an partial energy level diagram of 0.5 at. % $Pr^{3+}$ doped $Y_2SiO_5$ at a temperature of 5K. The pulse duration of DATA (D), WRITE (W), and READ (R) in Fig. 1a is set at 1 µs for the $\pi/2$ pulse area and 4.6 mW in power: Light A. The pulse duration of control pulse B1 (B2) is set at 1.2 µs for the $\pi$ ($3\pi$) pulse area and 14.5 mW in power: Light B. Lights R and B are used for initial preparation step to make $\rho_{11}=1$ before sending the pulse train in Fig. 1a. Figure 4b shows conventional three-pulse photon echoes as a function of $\Delta T$ (R delay from W) without using B1 and B2. As expected the measured decay time is 160 µs, which shows the optical population decay time $T_1^O$. The inset shows one of the data used for the plot, where the third peak is a two-pulse photon echo signal, and the last one is a stimulated photon echo signal: This last signal is measured for the plot. With the control pulses of B1 and B2, the measured decay time of 500 µs in Fig. 4c is the same as the spin homogeneous decay time $T_2^S$ measured by two-pulse Raman echoes (ref. 25). The two-pulse echo (3<sup>rd</sup> peak) in Fig. 4b is severely decreased by turning on the B1 control light, due to population transfer from the excited state $|3\rangle$ onto the auxiliary ground state $|2\rangle$. Because no perfect population transfer was made in the present scheme, the residual atoms cause coherence loss. So the optically locked echo intensity decreases to ~80% in comparison with the stimulated echo in Fig. 4a. Actually the delay $\Delta T$ also causes the coherence loss due to optical population decay. However, the loss percentage is negligibly small in the present time scale (ΔT=5 µs, ΔT<<$T_1^O$=160 µs). When the delay T of B2 increases beyond $T_2^S$, the decay process of echo signals shows saturation reaching a fixed value as shown in Fig. 4d (see red dashed line). For example, the echo intensity at t=2.5 ms decreases to ~75% in intensity, which is ~50% in amplitude. This means that the optically locked echo survives much longer than the conventional ultimate constraint of spin $T_2^S$ in most preexisting modified photon echo methods, satisfying more than 50% echo efficiency.

In conclusion, we proposed and demonstrated an optically locked photon echo protocol whose storage time is far beyond the limitation confined by most modified photon echo protocols whose storage time is determined by spin dephasing (spin inhomogeneous broadening). We measured the observed photon storage decay time follows the spin homogeneous decay for a while, but relaxed and eventually saturated at 50% echo efficiency even far beyond the ultimate limitation of spin homogeneous decay time. The present method is free from spontaneous emission noise controversy because of adapting three-pulse photon echo idea, not using a π pulse as the read-out process. The present method shows potential for ultralong quantum memory applications owing to extended storage time by orders of magnitude longer than in most modified photon echo methods. Ultralong photon storage is obtained via optical locking in conventional three-pulse photon echoes, where effectiveness of spin inhomogeneity is minimized.

**Acknowledgments**

This work was supported by the CRI program (Center for Photon Information Processing) of the Korean government (MEST) via National Research Foundation, S. Korea. BSH thanks R. W. Boyd of University of Rochester for helpful discussions.